\documentclass[12pt,twoside]{article}
\usepackage[mathscr]{eucal}
\usepackage{amsmath,amsfonts,amssymb,amsthm,mathabx,empheq}
\usepackage{times}
\usepackage{pdfsync}
\usepackage{cite}
\usepackage{url}
\usepackage{hyperref}
\usepackage{tensor}
\usepackage{color}
\usepackage{multicol}
\usepackage{bbold}

\advance\textheight\topskip
\allowdisplaybreaks[1]

\newcommand\td{\text{d}}
\newcommand\cO{{\cal O}}
\newcommand{\p}{\partial}

\newcommand{\be}{\begin{equation}}
\newcommand{\ee}{\end{equation}}
\newcommand{\bea}{\begin{eqnarray}}
\newcommand{\eea}{\end{eqnarray}}

\def\bz{\bar z}

\def\half{\frac12}

\def\bY{\bar Y}

\def\bp{\bar \partial}
\def\n{\nabla}
\def\bm{\bar{m}}
\def\bP{\bar P}
\def\bL{\bar{L}}

\def\bomega{\bar\omega}

\def\bL{\bar{L}}

\def \sd {\delta\hspace{-0.50em}\slash\hspace{-0.05em}}

\newcommand{\nn}{\nonumber}
\newcommand*\xbar[1]{%
  \hbox{%
    \vbox{%
      \hrule height 0.5pt 
      \kern0.3ex
      \hbox{%
        \kern-0.0em
        \ensuremath{#1}%
        \kern-0.0em
      }%
    }%
  }%
}

\allowdisplaybreaks[1]

\DeclareFontFamily{OT1}{rsfs}{} \DeclareFontShape{OT1}{rsfs}{m}{n}{
<-7> rsfs5 <7-10> rsfs7 <10-> rsfs10}{}
\DeclareMathAlphabet{\mycal}{OT1}{rsfs}{m}{n}

\hypersetup{
colorlinks=true,
linktoc=page,    
linkcolor=blue,
citecolor=blue,
urlcolor=blue,
colorlinks=true,
}

\begin{document}

\title{Gravitational vacua in the Newman-Penrose formalism}

\author{Pujian Mao}
\date{}

\def\mytitle{Gravitational vacua in the Newman-Penrose formalism}

\addtolength{\headsep}{4pt}

\begin{centering}

  \vspace{1cm}

  \textbf{\large{\mytitle}}

  \vspace{1cm}

  {\large Pujian Mao}

\vspace{0.5cm}

\begin{minipage}{.9\textwidth}\small \it  \begin{center}
     Center for Joint Quantum Studies and Department of Physics,\\
     School of Science, Tianjin University, 135 Yaguan Road, Tianjin 300350, China
 \end{center}
\end{minipage}

\end{centering}

\begin{center}
Emails:  pjmao@tju.edu.cn
\end{center}

\begin{center}
\begin{minipage}{.9\textwidth}
\textsc{Abstract}: In this paper, we derive the generic solution of the Newman-Penrose equations in the Newman-Unti gauge with vanishing curvature tensor. The obtained solutions are the vacua of the gravitational theory which are connected to the derivations in metric formalism from exponentiating the infinitesimal BMS generators in the BMS gauge in \cite{Compere:2016jwb,Compere:2016hzt} by a radial transformation. The coordinate transformations in the Newman-Unti gauge connecting each vacuum are also obtained. We confirm that the supertranslation charge of the gravitational vacua with respect to global considerations vanishes exactly not only in the Einstein theory but also when including the Holst, Pontryagin, and Gauss-Bonnet terms which verifies that the gravitational vacua are not affected by those trivial or boundary terms.

\end{minipage}
\end{center}

\thispagestyle{empty}


\section{Introduction}

A surprising result in the study of gravitational theory at the null infinity is the discovery of the BMS symmetry \cite{Bondi:1962px,Sachs:1962wk,Sachs:1962zza}. As asymptotic symmetry at the null infinity, it consists of an infinite dimensional supertranslation, which extends the Poincar\'{e} translations, and the Lorentz transformations. The physical role of the supertranslation at the perturbative level is in some sense a mystery until the recent proposal in \cite{Strominger:2013jfa} where it is argued that in a finite neighborhood of the Minkowski vacuum classical gravitational scattering and perturbative quantum gravity are BMS-invariant. The supertranslations are spontaneously broken in the conventional Minkowski vacuum, which
transform one vacuum solution to a new, physically inequivalent one. As a precise application of this proposal, it is proven that the supertranslation as broken symmetry controls the soft graviton theorem at the perturbative quantum level \cite{He:2014laa} and the gravitational memory effect at the classical level \cite{Strominger:2014pwa}, see also \cite{Strominger:2017zoo} for a comprehensive review.

The relevant investigations have been established mainly in the
asymptotic regions at null infinity. While the vacuum structure of the gravitational theory requires a detailed analysis anywhere of the spacetime in the bulk. This is also an urgent issue in the recent study of the emission of gravitational waves in the post-Minkowskian approach where some results have clear BMS observer dependence \cite{Damour:2020tta,Veneziano:2022zwh,Manohar:2022dea,Javadinezhad:2022ldc,Mao:2023evc,Riva:2023xxm}. In \cite{Compere:2016jwb,Compere:2016hzt}, the gravitational vacuum solutions are obtained by exponentiating the infinitesimal BMS generators in the bulk in the BMS gauge \cite{Bondi:1962px,Sachs:1962wk}. In this work, we will revisit the gravitational vacuum structure in the Newman-Penrose (NP) formalism \cite{Newman:1961qr} in the Newman-Unti (NU) gauge \cite{Newman:1962cia}. There are two reasons to study the gravitational vacua in the NP formalism in the NU gauge. First, the NP equations involve the full curvature tensor, while Einstein equations only involve the Ricci tensor. There is a clear way to derive the gravitational vacuum solutions from the NP equations by turning off the full curvature tensor. Technically, solving the NP equations is simpler than exponentiating the infinitesimal BMS generators. At least, NP equations are first order differential equations. But one needs to solve second order differential equations to deal with change of coordinates for the metric. Second, in the NU gauge, all the NP variables of the vacuum solutions are in terms of simple rational functions of the radial coordinate $r$. While square-root functions are involved in the BMS gauge \cite{Compere:2016jwb}, which arises from the  change of radial coordinate from the NU gauge to the BMS gauge.

In this paper, we derive the generic solutions of the NP equations with zero curvature tensor. The radial NP equations have a remarkable simplification where the solutions are given in terms of simple rational functions of $r$. The non-radial NP equations are exactly solvable in a divergence-free conformal frame \cite{Ashtekar:2014zsa}. The exact solutions of the NP equations consist of a supertranslation field, a superrotation field and a time-independent Weyl rescaling field. The precise coordinate transformations connecting each such vacua are also derived. First order formalism can contain new boundary degree of freedom which can not be seen from metric formalism, in particular when including trivial or boundary terms \cite{Godazgar:2018qpq,Godazgar:2020gqd,Godazgar:2020kqd,Godazgar:2022foc,Liu:2022uox,Mao:2022ldv}. This is another motivation to study the vacuum structure in the NP formalism where the spin coefficients of the vacuum solutions will be determined as well. Global requirements of asymptotic flatness \cite{Christodoulou:1993uv} will further reduce the gravitational vacua by turning off the superrotation field and fixing the Weyl factor. We compute the supertranslation charge of the reduced gravitational vacua from the Palatini, Holst, Pontryagin, and Gauss-Bonnet terms. We confirm that the supertranslation charges of the vacua from all the four terms are exactly zero when evaluated at any co-dimension two surface.

The organization of this paper is very simple. In section \ref{NP}, we briefly review the NP formalism and the NU gauge. In sectin \ref{solution}, we compute the most generic solution for the gravitational vacua in the divergence-free conformal frame. In section \ref{secplane} and \ref{secsphere}, we consider two particular boundary configurations, namely a complex plane and a sphere. The precise coordinate transformations relating each vacuum are derived. The supertranslation charge is computed. We conclude in the last section.

\section{NP formalism and NU gauge}
\label{NP}

The NP formalism \cite{Newman:1961qr} is a special tetrad formalism based on two real null basis, denoted as $e_1=l=e^2,\;e_2=n=e^1$, and two complex null basis $\;e_3=m=-e^4,\;e_4=\bar{m}=-e^3$, which are conjugate of each other to guarantee a Lorentzian spacetime. The four null basis vectors are orthogonalized and normalized as follows,
\be\label{tetradcondition}
l\cdot m=l\cdot\bm=n\cdot m=n\cdot\bm=0,\quad l\cdot n=1,\quad m\cdot\bm=-1.
\ee
The spacetime metric is determined from the tetrad as
\be
g_{\mu\nu}=n_\mu l_\nu + l_\mu n_\nu - m_\mu {\bm}_\nu - m_\nu \bm_\mu.
\ee
The spin connection in the NP formalism is labeled by 12 complex scalars denoted by special Greek symbols,
\be\label{coefficient}
\begin{split}
&\kappa=\Gamma_{311}=l^\nu m^\mu\nabla_\nu l_\mu,\;\;\pi=-\Gamma_{421}=-l^\nu \bar{m}^\mu\nabla_\nu n_\mu,\\
&\epsilon=\half(\Gamma_{211}-\Gamma_{431})=\half(l^\nu n^\mu\nabla_\nu l_\mu - l^\nu \bar{m}^\mu\nabla_\nu m_\mu),\\
&\tau=\Gamma_{312}=n^\nu m^\mu\nabla_\nu l_\mu,\;\;\nu=-\Gamma_{422}=-n^\nu \bar{m}^\mu\nabla_\nu n_\mu,\\
&\gamma=\half(\Gamma_{212}-\Gamma_{432})=\half(n^\nu n^\mu\nabla_\nu l_\mu - n^\nu \bar{m}^\mu\nabla_\nu m_\mu),\\
&\sigma=\Gamma_{313}=m^\nu m^\mu\nabla_\nu l_\mu,\;\;\mu=-\Gamma_{423}=-m^\nu \bar{m}^\mu\nabla_\nu n_\mu,\\
&\beta=\half(\Gamma_{213}-\Gamma_{433})=\half(m^\nu n^\mu\nabla_\nu l_\mu - m^\nu \bar{m}^\mu\nabla_\nu m_\mu),\\
&\rho=\Gamma_{314}=\bar{m}^\nu m^\mu\nabla_\nu l_\mu,\;\;\lambda=-\Gamma_{424}=-\bar{m}^\nu \bar{m}^\mu\nabla_\nu n_\mu,\\
&\alpha=\half(\Gamma_{214}-\Gamma_{434})=\half(\bar{m}^\nu n^\mu\nabla_\nu l_\mu - \bar{m}^\nu \bar{m}^\mu\nabla_\nu m_\mu).
\end{split}
\ee
We introduce the directional derivatives which are associated to the basis vectors as
\begin{align}
D=l^\mu\p_\mu,\;\;\;\;\Delta=n^\mu\p_\mu,\;\;\;\;\delta=m^\mu\p_\mu.
\end{align}
We would refer to \cite{Chandrasekhar}, also \cite{Liu:2022uox,Mao:2023yle,Mao:2024jpt}, for the other notations. We are only interested in the gravitational vacuum solutions in this work, hence the curvature tensor is zero.

The NU gauge \cite{Newman:1962cia} is well appreciated in the asymptotic analysis near the null infinity. This gauge is based on the existence of a hypersurface orthogonal null direction which is also tangent to null geodesics. Normally, this special null vector is chosen as basis $l$. Applying directly the orthogonality conditions and normalization conditions in \eqref{tetradcondition}, one can obtain the following relations from the definitions of the spin connection,
\be
\begin{split}\label{1}
&l^\nu \n_\nu l_\mu=(\epsilon+\bar \epsilon)l_\mu -\kappa \bm_\mu - \bar \kappa m_\mu,\\
&n^\nu \n_\nu l_\mu=(\gamma+\bar \gamma)l_\mu -\tau \bm_\mu - \bar \tau m_\mu,\\
&m^\nu \n_\nu l_\mu=(\beta+\bar \alpha)l_\mu -\sigma \bm_\mu - \bar \rho m_\mu,\\
&\bm^\nu \n_\nu l_\mu=(\alpha + \bar \beta)l_\mu - \rho \bm_\mu -\bar\sigma m_\mu,
\end{split}
\ee
\be
\begin{split}\label{2}
&l^\nu \n_\nu n_\mu=-(\epsilon+\bar \epsilon)n_\mu + \bar\pi \bm_\mu + \pi m_\mu,\\
&n^\nu \n_\nu n_\mu=-(\gamma+\bar \gamma)n_\mu + \bar\nu \bm_\mu + \nu m_\mu,\\
&m^\nu \n_\nu n_\mu=-(\beta+\bar \alpha)n_\mu + \bar\lambda \bm_\mu + \mu m_\mu,\\
&\bm^\nu \n_\nu n_\mu=-(\alpha + \bar \beta)n_\mu + \bar\mu \bm_\mu + \lambda m_\mu,\\
\end{split}
\ee
\be
\begin{split}\label{3}
&l^\nu \n_\nu m_\mu=(\epsilon - \bar\epsilon) m_\mu - \kappa n_\mu + \bar\pi l_\mu,\\
&n^\nu \n_\nu m_\mu=(\gamma - \bar\gamma) m_\mu  - \tau n_\mu + \bar\nu l_\mu,\\
&m^\nu \n_\nu m_\mu= (\beta - \bar\alpha) m_\mu - \sigma n_\mu + \bar\lambda l_\mu,\\
&\bm^\nu \n_\nu m_\mu= (\alpha - \bar\beta) m_\mu - \rho n_\mu + \bar\mu l_\mu,
\end{split}
\ee
Clearly, $l$ being tangent to null geodesics yields that $\kappa=0$ from the first equation in \eqref{1}. Then, one can use third and first classes of tetrad rotations to set $\epsilon=0$ and $\pi=0$. Consequently, $l$ is tangent to null geodesics with affine parameter and the rest three null basis are parallelly transported along $l$, which can be seen directly from the first equations in \eqref{2} and \eqref{3}. We define the tensor $B_{\nu\mu}^l=\n_\nu l_\mu$ to measure the geodesic deviation. After setting $\epsilon=\kappa=\pi=0$, the geodesic deviation $B_{\nu\mu}^l$ can be derived from the orthogonality conditions and normalization conditions of the tetrad vectors and the relations in \eqref{1} as
\begin{multline}\label{Bl}
B_{\nu\mu}^l=(\gamma+\bar \gamma)l_\mu l_\nu - \tau \bm_\mu l_\nu - \bar \tau m_\mu l_\nu - (\beta+\bar \alpha)l_\mu \bm_\nu - (\bar \beta+\alpha)l_\mu m_\nu \\ + \sigma \bm_\mu \bm_\nu + \bar \rho m_\mu \bm_\nu + \bar\sigma m_\mu m_\nu + \rho \bm_\mu m_\nu.
\end{multline}
Following the standard analysis of geodesic congruence \cite{Townsend:1997ku}, we can use the other null basis $-n$ to construct the transverse part of the metric $h_{\mu\nu}=g_{\mu\nu} - n_\mu l_\nu - l_\mu n_\nu$. Hence, $h_{\mu\nu}=- m_\mu {\bm}_\nu - m_\nu \bm_\mu$.
The transverse part of the geodesic deviation is
\be\label{Bl0}
\begin{split}
\hat{B}_{\nu\mu}^l=h^\alpha_\nu B_{\alpha\beta} h^\beta_\mu&=\sigma \bm_\mu \bm_\nu + \bar \rho m_\mu \bm_\nu + \bar\sigma m_\mu m_\nu + \rho \bm_\mu m_\nu\\
&=\sigma \bm_\mu \bm_\nu + \bar\sigma m_\mu m_\nu + (\bar \rho + \rho ) m_{(\mu} \bm_{\nu)} +  (\bar \rho - \rho ) m_{[\mu }\bm_{\nu]},
\end{split}
\ee
which manifests that $\rho-\bar\rho$ is the twist and $\rho=\bar\rho$ means $l$ is hypersurface orthogonal. For constructing a coordinate system, one normally choose the affine parameter as the radial coordinate $r$, hence $l=\frac{\p}{\p r}$. Since $l$ is hypersurface orthogonal, it must be proportional to the gradient of a scalar field. This scalar field can be chosen as a time coordinate $u$, hence $l=e^W du$. One can use the residual third class of tetrad rotation with a combined change of radial coordinate to set $W=\cO(r^{-1})$ without turning on $\epsilon$. Then, the commutation relation of the tetrad basis,
\be
D\Delta - \Delta D= - (\gamma + \xbar\gamma) D + \xbar\tau \delta + \tau \xbar\delta,
\ee
yields that $W=0$. Hence, $l$ is the gradient of a scalar field and the previous third class of tetrad rotation sets $\tau=\xbar\alpha+\beta$, which guarantees that $\n_\nu l_\mu + \n_\mu l_\nu=0$. This completes the NU gauge. One can use the usual angular variables $(\theta,\phi)$ for the rest two coordinates. Here, we take the stereographic coordinates $A=(z,\bz)$ which are
related to the usual angular variables by $z=e^{i\phi}\cot\frac{\theta}{2}$. Then, the tetrad and the co-tetrad must have the forms
\be
\begin{split}
&n=\frac{\p}{\p u} + U \frac{\p}{\p r} + X^A \frac{\p}{\p x^A},\\
&\label{gauge}l=\frac{\p}{\p r},\;\;\;\;\;\;m=\omega\frac{\p}{\p r} + L^A \frac{\p}{\p x^A},
\end{split}
\ee
and
\be
\begin{split}
&n= \left[-U-X^A(\xbar\omega L_A+\omega \bar L_A)\right]du + dr + (\omega\bar L_A+\xbar\omega L_A)dx^A,\\
&l=du,\;\;\;\;\;\;\;m=- X^AL_A du + L_A dx^A,
\end{split}
\ee
to satisfy the orthogonality and normalization conditions, where $L_AL^A=0,\;L_A\bar L^A=-1$. We will use $\p$ and $\bp$ to denote $\frac{\p}{\p z}$ and $\frac{\p}{\p \bz}$ for notational brevity.

\section{Vacuum solution}
\label{solution}

The vacuum of the gravitational theory is defined by the vanishing of curvature tensor. This will lead to a remarkable simplification in the solution space of the NP formalism where the $r$-dependence of all NP variables is in terms of simple rational functions.
The crucial step is to find the solutions for the following three radial NP equations
\begin{align}
\p_r \rho =\rho^2+\sigma\xbar\sigma,\quad\quad\quad\quad
\p_r \sigma=2 \rho \sigma,\quad\quad\quad\quad  \p_r \xbar\sigma=2 \rho \xbar\sigma.
\end{align}
Since $\rho$ is real imposed as a gauge condition, one can show that $\xbar\sigma=\sigma e^{i\Phi}$, where $\Phi$ is an arbitrary function independent of $r$, and $\rho=\frac12 \frac{\p_r \sigma}{\sigma}=\frac12 \frac{\p_r \xbar\sigma}{\xbar\sigma}$ in the case of $\sigma\neq0$ and $\xbar\sigma\neq0$. Inserting those relations to the first equation, one obtains
\be
\sigma \p_r^2 \sigma - \frac32  (\p_r \sigma)^2 - \sigma^4 e^{i\Phi}=0
\ee
and
\be
\xbar\sigma \p_r^2 \xbar\sigma - \frac32  (\p_r \xbar\sigma)^2 - \xbar\sigma^4 e^{-i\Phi}=0.
\ee
The generic solution to those equations are
\be
\sigma=\frac{\sigma^0}{r^2-\sigma^0\xbar\sigma^0},\quad\quad \xbar\sigma=\frac{\xbar\sigma^0}{r^2-\sigma^0\xbar\sigma^0},
\ee
where $\sigma^0$ and $\xbar\sigma^0$ are integration constants from the differential equations of $r$, namely two arbitrary functions independent of $r$. Then, the solution to $\rho$ is
\be
\rho=-\frac{r}{r^2-\sigma^0\xbar\sigma^0}
\ee
The rest radial NP equations can be solved in a similar way. The full solution to the radial equations are summarized as 
\begin{align}\label{Min}
&\rho=-\frac{r}{r^2-\sigma^0\xbar\sigma^0},\quad \sigma=\frac{\sigma^0}{r^2-\sigma^0\xbar\sigma^0},\quad \alpha=\frac{\alpha^0 r + \xbar\alpha^0 \xbar\sigma^0}{r^2-\sigma^0\xbar\sigma^0},\quad \beta=-\frac{\xbar\alpha^0 r + \alpha^0 \sigma^0}{r^2-\sigma^0\xbar\sigma^0},\nn \\
&L^z=-\frac{\sigma^0 \bP}{r^2-\sigma^0\xbar\sigma^0},\quad L^{\bz}=\frac{r P}{r^2-\sigma^0\xbar\sigma^0},\quad \omega= \frac{\omega^0 r - \xbar\omega^0 \sigma^0}{r^2-\sigma^0\xbar\sigma^0},\\
&\lambda=\frac{\lambda^0 r - \mu^0 \xbar\sigma^0}{r^2-\sigma^0\xbar\sigma^0},\quad \mu= \frac{\mu^0 r - \lambda^0 \sigma^0}{r^2-\sigma^0\xbar\sigma^0},\quad \gamma=\gamma^0,\quad \nu=\nu^0,\quad U=-(\gamma^0+\xbar\gamma^0)r + U^0,\nn
\end{align}
where $P$, $\bP$, and the quantities with superscript ``0'' are integration constants from the differential equations of $r$. The unlisted NP variables are zero. The integration constants are controlled by the non-radial NP equations as
\begin{align}
&\alpha^0=\half \bP \p \ln P,\quad\quad \gamma^0=-\half \p_u \ln \bP,\quad \quad\lambda^0= \p_u{\xbar\sigma^0} + \xbar \sigma^0 (3\gamma^0 - \xbar \gamma^0),\nn\\
&\mu^0= - \eth \alpha^0 - \xbar\eth\xbar\alpha^0=-\half P \bP \p\xbar\p \ln P\bP,\quad \quad \nu^0=\xbar \eth (\gamma^0+\xbar\gamma^0),\\
&\omega^0=\xbar\eth \sigma^0,\quad \quad U^0=\mu^0,\nn
\end{align}
where the operator $\eth$ for a field of spin $s$ is defined as,
\be
\eth \eta^{(s)}=(P\bp + 2 s \xbar\alpha^0)\eta^{(s)}=P\bP^{-s}\bp (\bP^s \eta^{(s)}),
\ee
and
\be
\xbar\eth \eta^{(s)}=(\bP\p  - 2 s \alpha^0)\eta^{(s)}=\bP P^{s}\p (P^{-s} \eta^{(s)}).
\ee
There are five equations constraining $P$, $\bP$, $\sigma^0$, and $\xbar\sigma^0$ as
\be\label{constraint}
\begin{split}
& \xbar\eth^2\sigma^0 -\eth^2\xbar\sigma^0 + \xbar\sigma^0\xbar\lambda^0 -\sigma^0\lambda^0=0,\quad \quad \eth\xbar \nu^0 - \p_u \xbar\lambda^0 - 4 \xbar\gamma^0 \xbar\lambda^0=0,\\
&\xbar \eth \mu^0 - \eth \lambda^0=0,\quad \quad  \eth \mu^0 - \xbar\eth \xbar\lambda^0=0,\quad \quad \xbar\eth \nu^0 - \p_u\lambda^0 - 4 \gamma^0 \lambda^0=0.
\end{split}
\ee
A solution of the above constraints will uniquely determine a solution of the NP equations which is one vacuum of the gravitational theory. The commutation relation of the $\eth$ operators is
\be\label{commutator}
[\xbar\eth,\eth]\eta^{(s)} =-2s\mu^0\eta^{(s)},
\ee
which is useful for deriving the solution to equations in \eqref{constraint}. The spin weights of relevant fields are listed in Table \ref{t1}.
\begin{table}[ht]
\caption{Spin weights}\label{t1}
\begin{center}\begin{tabular}{|c|c|c|c|c|c|c|c|c|c|c|c|c|c|c|c|c|c|c|c|c}
\hline
& $\eth$ & $\omega^0$ & $\alpha^0$ & $\gamma^0$ & $\nu^0$ & $\mu^0$ & $\sigma^0$ & $\lambda^0$    \\
\hline
s & $1$& $1$& $-1$ &$0$& $-1$& $0$& $2$& $-2$     \\
\hline
\end{tabular}\end{center}\end{table}

For any $u$-independent $P(z,\bz)$ and $\bP(z,\bz)$, the constraints in \eqref{constraint} are exactly solvable. Such conditions lead to $\gamma^0=0=\nu^0$ which define a divergence-free conformal frame \cite{Ashtekar:2014zsa} and have clear geometric meaning from the following relation in \eqref{2},
\be
n^\nu \n_\nu n_\mu=-(\gamma+\bar \gamma)n_\mu + \bar\nu \bm_\mu + \nu m_\mu.
\ee
It means that $n$ is tangent to null geodesic at null infinity with affine parameter, namely $n$ is the generator of the null infinity. Since $n$ is simply $\frac{\p}{\p u}$ at the null infinity in this case, $u$ is the affine parameter. Hence $u\in(-\infty,+\infty)$ guarantees the geodesic completeness of the null infinity. The solution to the equations in \eqref{constraint} with such conditions is
\be\label{sigma}
\begin{split}
&\sigma^0=\left[u-C(z,\bz)\right]\xbar\lambda^0(z,\bz)  + \eth^2 C(z,\bz),\\
&\xbar\sigma^0=\left[u-C(z,\bz)\right] \lambda^0(z,\bz) + \xbar\eth^2 C(z,\bz),
\end{split}
\ee
and
\be\label{lambda}
\begin{split}
&\lambda^0=\bP^2 \Lambda(z) + \frac{1}{4P^2}\left[(P \p \bP - \bP \p P)^2-2P \bP (\bP \p^2 P + P \p^2 \bP)\right] ,\\
&\xbar\lambda^0= P^2 \xbar\Lambda(\bz) + \frac{1}{4 \bP^2}\left[(\bP \bp P - P \bp \bP)^2-2P \bP (P \bp^2 \bP + \bP \bp^2 P)\right],
\end{split}
\ee
where $P(z,\bz)$ is completely free and $\bP(z,\bz)$ is its complex conjugate.
Comparing to the finite BMS transformation results in \cite{Compere:2016hzt,Compere:2016jwb,Barnich:2016lyg},
it is clear that $C(z,\bz)$ characterizes the supertranslation, $\Lambda(z)$ and $\xbar\Lambda(\bz)$ represent the superrotation (see also the infinitesimal studies \cite{Barnich:2009se,Barnich:2010eb}) and $\sqrt{P\bP}$ determines the Weyl rescaling of the two surface at null infinity at any constant $u$.\footnote{The functions $\Lambda(z)$ and $\xbar\Lambda(\bz) $ are given in the form of Schwarzian derivative of a holomorphic and an anti-holomorphic function in the finite transformation papers \cite{Compere:2016hzt,Compere:2016jwb,Barnich:2016lyg}.} The precise coordinate transformations from Minkowski vacuum in the common sense to the above solution will be presented in the next sections. The co-tetrad basis of the solutions are in the form
\be\label{co-tetrad}
\begin{split}
l=&du,\quad \quad n= -U^0 du + dr - \frac{\xbar\omega^0}{\bP} d z - \frac{\omega^0}{P} d \bz,\\
m=& - \frac{r}{\bP}dz - \frac{\sigma^0}{P} d\bz,\quad \quad \bm= - \frac{\xbar\sigma^0}{\bP} d z  - \frac{r}{P}d \bz ,
\end{split}
\ee
which yields the line-element as
\begin{multline}\label{generic}
ds^2= -2 U^0 du^2 + 2dudr  - \frac{ 2\xbar\omega^0}{\bP} du d z - \frac{2 \omega^0}{P} du d \bz  \\
-  \frac{2r\xbar\sigma^0}{\bP^2} dz^2 -  \frac{2r\sigma^0}{P^2} d\bz^2 -  \frac{2(r^2 + \sigma^0\xbar\sigma^0)}{P \bP}dz d\bz.
\end{multline}

The above solutions are obtained only with respect to local equations of motion. As a vacuum solution of gravitational theory, there are also global requirements such as the Christodoulou-Klainerman asymptotic flatness conditions \cite{Christodoulou:1993uv}, which requires that $\lambda^0$ decays as $|u|^{-\frac32}$ for large $u$. For the current case, such conditions yield $\lambda^0=0$. Note also that the solution of $\lambda^0$ in \eqref{lambda} implies the energy of the system unbounded from below \cite{Compere:2016jwb}. Asymptotic flatness conditions yield that $\mu^0$ is a constant, which implies that the two surface at null infinity at any constant $u$ must have constant curvature. We will investigate in details two common choices, namely a plane and a sphere, in the next sections.

\section{Complex plane boundary}
\label{secplane}

One particularly simple solution to a constant $\mu^0$ is $P=\bP=1$. Correspondingly, $\mu^0=0$. This choice fixes the boundary line-element as $\td s^2=2\td z \td\bz$.
Hence, the null infinity is with topology $\mathbb{R}\times \mathbb{C}$. The gravitational vacua are entirely spanned by supertranslation field $C$. The solution is
\be\label{Min1}
\sigma^0= \bp^2 C(z,\bz),\quad \quad \xbar\sigma^0= \p^2 C(z,\bz),
\ee
and
\be\label{Min2}
\alpha^0=0,\quad \gamma^0=0, \quad  U^0=\mu^0=0,\quad  \omega^0=\p \sigma^0, \quad \lambda^0=0, \quad \nu^0=0.
\ee
The line-element of the spacetime is
\begin{multline}\label{superplane}
ds^2= 2dudr  - 2 \p \p \bp C du d z - 2 \bp \p \bp C du d \bz  \\
- 2r \p^2 C dz^2 -  2r \bp^2 C d\bz^2 -  2(r^2 + \p^2 C \bp^2 C) dz d\bz.
\end{multline}
Turning off the supertranslation field $C$, one recovers the Minkowski spacetime in flat null coordinates,
\be\label{plane}
ds^2= 2d u_p d r_p
 - 2 r_p^2 d z_p d\bz_p.
\ee
Similar to the case in BMS gauge \cite{Compere:2016hzt}, the generic solution \eqref{generic} can be obtained from \eqref{plane} by the following change of coordinates,
\be\label{generalchange}
\begin{split}
&r_p=\frac{r}{\p_u \Phi}+\frac{\p\bp \Phi}{\p Y \bp \bY},\\
&u_p=\Phi-\frac{\p \Phi \bp \Phi}{\p Y \bp \bY r_p},\\
&z_p=Y(z)-\frac{\bp \Phi}{\bp \bY r_p},\\
&\bz_p=\bY(\bz)-\frac{\p \Phi}{\p Y r_p},
\end{split}
\ee
where $\Phi=\sqrt{P\bP} (u-C)\sqrt{ \p Y \bp \bY}$.\footnote{One can generalize the transformation to obtain a $u$-dependent $P$ and $\bP$ solution by setting $\Phi= (u-C)\sqrt{ \p Y \bp \bY}\int du' (\sqrt{P(u',z,\bz)\bP(u',z,\bz)}) $, which gives the most generic local vacuum solutions.} The superrotation field $\Lambda(z)$ in \eqref{lambda} is defined by $\Lambda(z)=-\frac12 S[Y(z),z]$ through the change of coordinates, where $S[Y(z),z]$ is the Schwarzian derivative of $Y(z)$,
\be
S[Y(z),z]=\frac{Y'''}{Y'}-\frac{3(Y'')^2}{2(Y')^2}.
\ee
Hence, turning off $\Lambda(z)$ reduces the superrotations to Lorentz transformations. The coordinate transformations in particular connect \eqref{plane} and the supertranslated one \eqref{superplane} are
\be
\begin{split}
&r_p=r-\p \bp C,\\
&u_p=u-C-\frac{\p C \bp C}{r_p},\\
&z_p=z+\frac{\bp C}{r_p}.
\end{split}
\ee
We will consider the solution in \eqref{superplane} as the gravitational vacua and compute the supertranslation charge of this solution space. It is proven \cite{Compere:2016jwb} that the supertranslation charge in the BMS gauge is exactly zero anywhere in the vacuum spacetime. Here, we will check the supertranslation charge when the theory is coupled with some trivial or boundary terms, including the Holst, Pontryagin, and Gauss-Bonnet terms. It is shown that the form language is particularly convenient for computing surface charges for first order formalism \cite{Godazgar:2020gqd,Godazgar:2020kqd}. The exact formulas of the forms $e_a$ are given in \eqref{co-tetrad} and the connection one forms $\Gamma_{ab}$ in the NU gauge are given as
\bea
&& \Gamma_{12}=-(\gamma + \xbar\gamma) l + \xbar\tau m  +\tau \bm,\\
&&\Gamma_{13}= -\tau l + \rho m + \sigma \bm,\\
&&\Gamma_{23}=\xbar\nu l - \xbar\mu m - \xbar\lambda \bm ,\\
&&\Gamma_{34}=(\gamma -\xbar\gamma) l - (\alpha-\xbar\beta) m  + (\xbar\alpha- \beta ) \bm.
\eea
With adaption to the solution \eqref{superplane}, the forms are given by
\be\label{e}
\begin{split}
e_1=&du,\quad \quad e_2= dr - \bp \p^2 C d z - \p \bp^2 C d \bz,\\
e_3=& - r dz - \bp^2 C d\bz,\quad \quad e_4= - \p^2 C d z  - r d \bz ,
\end{split}
\ee
and
\be\label{Gamma}
\Gamma_{12}=0,\quad \quad \Gamma_{13}= d z,\quad \quad \Gamma_{23}=0,\quad \quad \Gamma_{34}=0.
\ee
The gauge transformation of the NP formalism is a combination of a diffeomorphism and a local Lorentz transformation. The actions on the tetrad and spin connection are given by
\be
\begin{split}
&\delta_{\xi,\omega}{e_a}^\mu ={\xi}^\nu\partial_\nu
{e_a}^\mu-\p_\nu{\xi}^\mu{e_a}^\nu +{\omega_a}^b{e_b}^\mu, \\
&\delta_{\xi, \omega} \Gamma_{a b c} = {\xi}^\nu \partial_\nu \Gamma_{a b c} - e_c^\mu \p_\mu {\omega}_{a b} + {\omega_a}^{d}\Gamma_{dbc}+ {\omega_b}^{d}\Gamma_{adc} + {\omega_c}^{d}\Gamma_{abd} .
\end{split}
\ee
A supertranslation for a generic spacetime is obtained in \cite{Barnich:2019vzx},
\begin{equation}
\begin{split}
&\xi^u = T(z,\bz),\quad \quad
\xi^A= - \p_B T \int^{+\infty}_r dr[L^A \bL^B +  \bL^AL^B ], \\
&\xi^r= \xbar\eth
\eth T - \p_A T
\int^{+\infty}_r dr[\omega \bL^A + \bomega L^A + X^A], \\
\end{split}
\end{equation}
and
\begin{equation}
\begin{split}
\omega^{12}=&  X^A \p_A T, \quad
\omega^{23}= \bL^A \p_A T,\quad
\omega^{24}= L^A \p_A T, \\
\omega^{13}=&(\gamma^0 + \bar{\gamma}^0) \xbar\eth T  + \p_A T \int^{+\infty}_r dr[\lambda
L^A + \mu \bL^A], \\
\omega^{14}=&(\gamma^0 + \bar{\gamma}^0) \eth T  + \p_A T \int^{+\infty}_r dr[\bar{\lambda}
\bL^A + \bar{\mu} L^A], \\
\omega^{34}=&  - \partial_A T
\int^{+\infty}_{r} dr [(\bar{\alpha} - \beta ) \bar{L}^A +
(\bar{\beta} - \alpha) L^A].
\end{split}
\end{equation}
Inserting the NP variables of the supertranslated Minkowski spacetime \eqref{Min1} and \eqref{Min2}, the corresponding supertranslation is given by 
\be\begin{split}\label{xi}
&\xi^u=T(z,\bz),\quad \quad \xi^z= \left( \frac{\p T \sigma^0 }{r^2-\sigma^0\xbar\sigma^0} -  \frac{\bp T r}{r^2-\sigma^0\xbar\sigma^0} \right),\\
&\xi^r= \p\bp T - \frac{\p T (\omega^0 r - \xbar\omega^0 \sigma^0)}{r^2-\sigma^0\xbar\sigma^0}  -\frac{ \bp T (\xbar\omega^0 r - \omega^0 \xbar\sigma^0)}{r^2-\sigma^0\xbar\sigma^0} ,
\end{split}
\ee
and
\be\begin{split}\label{omega}
&\omega_{12}=0,\quad \quad \omega_{13}=\left( \frac{\p T \sigma^0 }{r^2-\sigma^0\xbar\sigma^0} -  \frac{\bp T r}{r^2-\sigma^0\xbar\sigma^0} \right),\\
&\omega_{23}= 0,\quad \quad \omega_{34}=0.
\end{split}
\ee
Note that $\sigma^0=\bp^2 C$. The surface charge with respect to the Palatini, Holst, Pontryagin, and Gauss-Bonnet terms are given by \cite{Godazgar:2020gqd,Godazgar:2020kqd,Godazgar:2022foc,Liu:2022uox,Mao:2022ldv}
\be\label{palatinicharge}
\sd {\cal H}_{Pa}= \frac{1}{32\pi G} \epsilon^{abcd} \int_{\partial \Sigma} \left[\delta (i_\xi \Gamma_{ab}  e_c \wedge e_d) - i_\xi( \delta \Gamma_{ab}\wedge e_c\wedge e_d) - \delta(\omega_{ab} e_c\wedge e_d)\right],
\ee
\be
\sd {\cal H}_H= \frac{it}{16\pi G}  \int_{\partial \Sigma} \left[\delta(i_\xi \Gamma^{ab}  e_a \wedge e_b) - i_\xi( \delta \Gamma^{ab}\wedge  e_a \wedge e_b) -\delta( \omega^{ab} e_a\wedge e_b)\right].
\ee
\be\label{Pcharge}
\sd {\cal H}_{Po}= \frac{1}{16\pi G}\int_{\partial \Sigma}
\delta \Gamma^{ab}\wedge \delta_{\xi,\omega} \Gamma_{ab}
\ee
\be\label{GBcharge}
\sd {\cal H}_{GB}= \frac{1}{16\pi G}\epsilon^{abcd}\int_{\partial \Sigma}
\delta \Gamma_{ab}\wedge \delta_{\xi,\omega} \Gamma_{cd},
\ee
respectively, where $\partial \Sigma$ can be any co-dimension two surface of the spacetime to evaluate the charge. By inserting the solutions in \eqref{e} and \eqref{Gamma}, and the symmetry parameters in \eqref{xi} and \eqref{omega} into the charge expressions, one can prove directly that all components of the supertranslation charge from those four terms are exactly zero. In particular, it is very easy to see that the charges from Pontryagin and Gauss-Bonnet terms are zero, because the components of the spin connection in \eqref{Gamma} are all given by definite functions, namely $\delta_{\xi,\omega} \Gamma_{ab}=0$.

\section{Sphere boundary}
\label{secsphere}

In this section, we will repeat what has been done in the previous section for another solution of constant $\mu^0$. The solution is given by $P=\bP=P_s=\frac{1+z\bz}{\sqrt2}$. This choice fixes the boundary line-element as $\td s^2=\frac{4}{(1+z\bz)^2}\td z \td\bz$, which is a unit sphere. Hence, the null infinity is with topology $\mathbb{R}\times \mathbb{S}^2$. The exact forms of the solution are
\be
\sigma^0= \eth^2 C(z,\bz),\quad \quad \xbar\sigma^0= \xbar\eth^2 C(z,\bz),
\ee
and
\be\label{spheresolution}
\alpha^0=\frac{\bz}{2\sqrt2},\quad \gamma^0=0, \quad  U^0=\mu^0=-\frac12,\quad  \omega^0=\xbar\eth \sigma^0, \quad \lambda^0=0, \quad \nu^0=0.
\ee
The line-element of this spacetime is
\begin{multline}\label{supersphere}
ds^2=du^2 + 2dudr  - \frac{ 2 \eth\xbar\eth^2 C}{P_s} du d z - \frac{2 \xbar\eth \eth^2 C}{P_s} du d \bz  \\
-  \frac{2r\xbar\eth^2 C}{P_s^2} dz^2 -  \frac{2r\eth^2 C}{P_s^2} d\bz^2 -  \frac{2(r^2 + \eth^2 C \xbar\eth^2 C)}{P_s^2 }dz d\bz.
\end{multline}
This solution is connected to the one in the BMS gauge in \cite{Compere:2016jwb} by a radial transformation $r=\sqrt{r_{CL}^2 + U}$, where $r_{CL}$ is the radial coordinate in \cite{Compere:2016jwb} and $U$ is defined in Eq. (2.4) of that reference. Note also that the signature of the NP formalism is $(+,+,+,-)$ which is different from the one in \cite{Compere:2016jwb}. Turning off the supertranslation field $C$ recovers the global Minkowski vacuum in retarded stereographic coordinates,
\be\label{sphere}
d s^2=d u_s^2+ 2d u_s d r_s
 - \frac{2r_s^2}{P_s^2} d z_s d\bz_s.
\ee
Following the strategy in \cite{Compere:2016hzt}, the coordinate transformations connecting the line-elements \eqref{sphere} and \eqref{supersphere} include two steps. First, one goes from the sphere boundary to the plane case in \eqref{plane} by
\be
\begin{split}
&r_s=\frac{1}{\sqrt2} \sqrt{[u_p + r_p(1+z_p \bz_p)]^2-4 r_p u_p},\\
&u_s=\frac{1}{\sqrt2} [u_p + r_p(1+z_p \bz_p)] - r_s,\\
&z_s=\frac{\frac{1}{\sqrt2} [u_p - r_p(1+z_p \bz_p)]+r_s}{u_s \bz_p}.
\end{split}
\ee
Then, we specify the generic change of coordinates in \eqref{generalchange} to a sphere case by
\be
\begin{split}
&r_p=\frac{r}{P_s}+\p\bp \Phi,\\
&u_p=\Phi-\frac{\p \Phi \bp \Phi}{r_p},\\
&z_p=z-\frac{\bp \Phi}{r_p},
\end{split}
\ee
where $\Phi=P_s (u-C)$ now.  The exact formulas of the tetrad form and the connection one form with respect to \eqref{supersphere} are given as
\be
\begin{split}
e_1=&du,\quad \quad e_2= \frac12 du + dr - \frac{\eth\xbar\eth^2 C}{P_s} d z - \frac{\xbar\eth\eth^2 C}{P_s} d \bz,\\
e_3=& - \frac{r}{P_s}dz - \frac{\eth^2 C}{P_s} d\bz,\quad \quad e_4= - \frac{\xbar\eth^2 C}{P_s} d z  - \frac{r}{P_s}d \bz ,
\end{split}
\ee
and
\be
\Gamma_{12}=0,\quad \quad \Gamma_{13}= \frac{1}{P_s} dz,\quad \quad \Gamma_{23}= \frac{\mu^0}{P_s} dz,\quad \quad \Gamma_{34}= \frac{2}{P_s} \left(\alpha^0 dz - \xbar \alpha^0 d \bz\right).
\ee
Note that $\mu^0$ and $\alpha^0$ should be replaced by the expressions in \eqref{spheresolution} in the beginning of this section. A supertranslation in such case is
\be\begin{split}
&\xi^u=T(z,\bz),\quad \quad \xi^z= P_s\left( \frac{\xbar\eth T \sigma^0 }{r^2-\sigma^0\xbar\sigma^0} -  \frac{\eth T r}{r^2-\sigma^0\xbar\sigma^0} \right),\\\nn
&\xi^r=P_s^2 \p\bp T - \frac{\xbar\eth T (\omega^0 r - \xbar\omega^0 \sigma^0)}{r^2-\sigma^0\xbar\sigma^0}  -\frac{ \eth T (\xbar\omega^0 r - \omega^0 \xbar\sigma^0)}{r^2-\sigma^0\xbar\sigma^0} ,
\end{split}
\ee
and
\be\begin{split}\nn
&\omega_{12}=0,\quad \quad \omega_{13}=\left( \frac{\xbar\eth T \sigma^0 }{r^2-\sigma^0\xbar\sigma^0} -  \frac{\eth T r}{r^2-\sigma^0\xbar\sigma^0} \right),\\
&\omega_{23}= \mu^0 \omega_{13},\quad \quad \omega_{34}=\frac{2 \xbar\eth T (\xbar\alpha^0 r + \alpha^0 \sigma^0)}{r^2-\sigma^0\xbar\sigma^0}  -\frac{ 2 \eth T (\alpha^0 r + \xbar\alpha^0 \xbar\sigma^0)}{r^2-\sigma^0\xbar\sigma^0}.
\end{split}
\ee
Note that $\sigma^0=\eth^2 C$ now. We verify that all components of the supertranslation charge from Palatini, Holst, Pontryagin, and Gauss-Bonnet terms are exactly zero.

\section{Conclusion}

To conclude, we study the vacuum structure of the gravitational theory. The vacua are obtained in a self-contained way from the NP equations in the NU gauge. The transformations connecting each vacuum solution are revealed. We show that the supertranslation charges of the gravitational vacua with respect to global considerations from the Palatini, Holst, Pontryagin, and Gauss-Bonnet terms are exactly zero anywhere in the spacetime.

\section*{Acknowledgments}

This work is supported in part by the National Natural Science Foundation of China (NSFC) under Grants No. 11935009 and No. 11905156.

\providecommand{\href}[2]{#2}\begingroup\raggedright\endgroup


\begin{thebibliography}{10}

\bibitem{Compere:2016jwb}
G.~Comp\`ere and J.~Long, ``{Vacua of the gravitational field},''
  \href{http://dx.doi.org/10.1007/JHEP07(2016)137}{{\em JHEP} {\bfseries 07}
  (2016) 137}, \href{http://arxiv.org/abs/1601.04958}{{\ttfamily
  arXiv:1601.04958 [hep-th]}}.

\bibitem{Compere:2016hzt}
G.~Comp\`ere and J.~Long, ``{Classical static final state of collapse with
  supertranslation memory},''
  \href{http://dx.doi.org/10.1088/0264-9381/33/19/195001}{{\em Class. Quant.
  Grav.} {\bfseries 33} no.~19, (2016) 195001},
  \href{http://arxiv.org/abs/1602.05197}{{\ttfamily arXiv:1602.05197 [gr-qc]}}.

\bibitem{Bondi:1962px}
H.~Bondi, M.~G.~J. van~der Burg, and A.~W.~K. Metzner, ``{Gravitational waves
  in general relativity. 7. Waves from axisymmetric isolated systems},''
  \href{http://dx.doi.org/10.1098/rspa.1962.0161}{{\em Proc. Roy. Soc. Lond. A}
  {\bfseries 269} (1962) 21--52}.

\bibitem{Sachs:1962wk}
R.~K. Sachs, ``{Gravitational waves in general relativity. 8. Waves in
  asymptotically flat space-times},''
  \href{http://dx.doi.org/10.1098/rspa.1962.0206}{{\em Proc. Roy. Soc. Lond. A}
  {\bfseries 270} (1962) 103--126}.

\bibitem{Sachs:1962zza}
R.~Sachs, ``{Asymptotic symmetries in gravitational theory},''
  \href{http://dx.doi.org/10.1103/PhysRev.128.2851}{{\em Phys. Rev.} {\bfseries
  128} (1962) 2851--2864}.

\bibitem{Strominger:2013jfa}
A.~Strominger, ``{On BMS Invariance of Gravitational Scattering},''
  \href{http://dx.doi.org/10.1007/JHEP07(2014)152}{{\em JHEP} {\bfseries 07}
  (2014) 152}, \href{http://arxiv.org/abs/1312.2229}{{\ttfamily arXiv:1312.2229
  [hep-th]}}.

\bibitem{He:2014laa}
T.~He, V.~Lysov, P.~Mitra, and A.~Strominger, ``{BMS supertranslations and
  Weinberg\textquoteright{}s soft graviton theorem},''
  \href{http://dx.doi.org/10.1007/JHEP05(2015)151}{{\em JHEP} {\bfseries 05}
  (2015) 151}, \href{http://arxiv.org/abs/1401.7026}{{\ttfamily arXiv:1401.7026
  [hep-th]}}.

\bibitem{Strominger:2014pwa}
A.~Strominger and A.~Zhiboedov, ``{Gravitational Memory, BMS Supertranslations
  and Soft Theorems},'' \href{http://dx.doi.org/10.1007/JHEP01(2016)086}{{\em
  JHEP} {\bfseries 01} (2016) 086},
  \href{http://arxiv.org/abs/1411.5745}{{\ttfamily arXiv:1411.5745 [hep-th]}}.

\bibitem{Strominger:2017zoo}
A.~Strominger, {\em {Lectures on the Infrared Structure of Gravity and Gauge
  Theory}}.
\newblock Princeton University Press, Princeton, 2018.
\newblock \href{http://arxiv.org/abs/1703.05448}{{\ttfamily arXiv:1703.05448
  [hep-th]}}.

\bibitem{Damour:2020tta}
T.~Damour, ``{Radiative contribution to classical gravitational scattering at
  the third order in $G$},''
  \href{http://dx.doi.org/10.1103/PhysRevD.102.124008}{{\em Phys. Rev. D}
  {\bfseries 102} no.~12, (2020) 124008},
  \href{http://arxiv.org/abs/2010.01641}{{\ttfamily arXiv:2010.01641 [gr-qc]}}.

\bibitem{Veneziano:2022zwh}
G.~Veneziano and G.~A. Vilkovisky, ``{Angular momentum loss in gravitational
  scattering, radiation reaction, and the Bondi gauge ambiguity},''
  \href{http://dx.doi.org/10.1016/j.physletb.2022.137419}{{\em Phys. Lett. B}
  {\bfseries 834} (2022) 137419},
  \href{http://arxiv.org/abs/2201.11607}{{\ttfamily arXiv:2201.11607 [gr-qc]}}.

\bibitem{Manohar:2022dea}
A.~V. Manohar, A.~K. Ridgway, and C.-H. Shen, ``{Radiated Angular Momentum and
  Dissipative Effects in Classical Scattering},''
  \href{http://dx.doi.org/10.1103/PhysRevLett.129.121601}{{\em Phys. Rev.
  Lett.} {\bfseries 129} no.~12, (2022) 121601},
  \href{http://arxiv.org/abs/2203.04283}{{\ttfamily arXiv:2203.04283
  [hep-th]}}.

\bibitem{Javadinezhad:2022ldc}
R.~Javadinezhad and M.~Porrati, ``{Supertranslation-Invariant Formula for the
  Angular Momentum Flux in Gravitational Scattering},''
  \href{http://dx.doi.org/10.1103/PhysRevLett.130.011401}{{\em Phys. Rev.
  Lett.} {\bfseries 130} no.~1, (2023) 011401},
  \href{http://arxiv.org/abs/2211.06538}{{\ttfamily arXiv:2211.06538 [gr-qc]}}.

\bibitem{Mao:2023evc}
P.~Mao, J.-B. Wu, and X.~Wu, ``{Angular momentum and memory effect},''
  \href{http://dx.doi.org/10.1103/PhysRevD.107.L101501}{{\em Phys. Rev. D}
  {\bfseries 107} no.~10, (2023) L101501},
  \href{http://arxiv.org/abs/2301.08032}{{\ttfamily arXiv:2301.08032 [gr-qc]}}.

\bibitem{Riva:2023xxm}
M.~M. Riva, F.~Vernizzi, and L.~K. Wong, ``{Angular momentum balance in
  gravitational two-body scattering: Flux, memory, and supertranslation
  invariance},'' \href{http://dx.doi.org/10.1103/PhysRevD.108.104052}{{\em
  Phys. Rev. D} {\bfseries 108} no.~10, (2023) 104052},
  \href{http://arxiv.org/abs/2302.09065}{{\ttfamily arXiv:2302.09065 [gr-qc]}}.

\bibitem{Newman:1961qr}
E.~Newman and R.~Penrose, ``{An Approach to gravitational radiation by a method
  of spin coefficients},''
\href{http://dx.doi.org/10.1063/1.1724257}{{\em J. Math. Phys.} {\bfseries 3}
  (1962) 566--578}.

\bibitem{Newman:1962cia}
E.~T. Newman and T.~W.~J. Unti, ``{Behavior of Asymptotically Flat Empty
  Spaces},'' \href{http://dx.doi.org/10.1063/1.1724303}{{\em J. Math. Phys.}
  {\bfseries 3} no.~5, (1962) 891}.

\bibitem{Ashtekar:2014zsa}
A.~Ashtekar, ``{Geometry and Physics of Null Infinity},''
  \href{http://arxiv.org/abs/1409.1800}{{\ttfamily arXiv:1409.1800 [gr-qc]}}.

\bibitem{Godazgar:2018qpq}
H.~Godazgar, M.~Godazgar, and C.~N. Pope, ``{New dual gravitational charges},''
  \href{http://dx.doi.org/10.1103/PhysRevD.99.024013}{{\em Phys. Rev. D}
  {\bfseries 99} no.~2, (2019) 024013},
  \href{http://arxiv.org/abs/1812.01641}{{\ttfamily arXiv:1812.01641
  [hep-th]}}.

\bibitem{Godazgar:2020gqd}
H.~Godazgar, M.~Godazgar, and M.~J. Perry, ``{Asymptotic gravitational
  charges},'' \href{http://dx.doi.org/10.1103/PhysRevLett.125.101301}{{\em
  Phys. Rev. Lett.} {\bfseries 125} no.~10, (2020) 101301},
  \href{http://arxiv.org/abs/2007.01257}{{\ttfamily arXiv:2007.01257
  [hep-th]}}.

\bibitem{Godazgar:2020kqd}
H.~Godazgar, M.~Godazgar, and M.~J. Perry, ``{Hamiltonian derivation of dual
  gravitational charges},''
  \href{http://dx.doi.org/10.1007/JHEP09(2020)084}{{\em JHEP} {\bfseries 09}
  (2020) 084}, \href{http://arxiv.org/abs/2007.07144}{{\ttfamily
  arXiv:2007.07144 [hep-th]}}.

\bibitem{Godazgar:2022foc}
M.~Godazgar, G.~Macaulay, and G.~Long, ``{Higher derivative asymptotic charges
  and internal Lorentz symmetries},''
  \href{http://dx.doi.org/10.1103/PhysRevD.105.084037}{{\em Phys. Rev. D}
  {\bfseries 105} no.~8, (2022) 084037},
  \href{http://arxiv.org/abs/2201.07014}{{\ttfamily arXiv:2201.07014
  [hep-th]}}.

\bibitem{Liu:2022uox}
H.-S. Liu and P.~Mao, ``{Near horizon gravitational charges},''
  \href{http://dx.doi.org/10.1007/JHEP05(2022)123}{{\em JHEP} {\bfseries 05}
  (2022) 123}, \href{http://arxiv.org/abs/2201.10308}{{\ttfamily
  arXiv:2201.10308 [hep-th]}}.

\bibitem{Mao:2022ldv}
P.~Mao and W.~Zhao, ``{Null boundary gravitational charges from local Lorentz
  symmetries},'' \href{http://dx.doi.org/10.1103/PhysRevD.107.044004}{{\em
  Phys. Rev. D} {\bfseries 107} no.~4, (2023) 044004},
  \href{http://arxiv.org/abs/2211.04736}{{\ttfamily arXiv:2211.04736
  [hep-th]}}.

\bibitem{Christodoulou:1993uv}
D.~Christodoulou and S.~Klainerman, {\em {The Global nonlinear stability of the
  Minkowski space}}.
\newblock Princeton University Press, Princeton, 1993.

\bibitem{Chandrasekhar}
S.~Chandrasekhar, ``{The Newman-Penrose formalism},'' in {\em {The mathematical
  theory of black holes}}, ch.~1, pp.~40--55.
\newblock Oxford, UK, 1983.

\bibitem{Mao:2023yle}
P.~Mao and W.~Zhao, ``{Asymptotic Weyl double copy in Newman-Penrose
  formalism},'' \href{http://dx.doi.org/10.1007/JHEP02(2024)171}{{\em JHEP}
  {\bfseries 02} (2024) 171}, \href{http://arxiv.org/abs/2312.17160}{{\ttfamily
  arXiv:2312.17160 [hep-th]}}.

\bibitem{Mao:2024jpt}
P.~Mao and W.~Zhao, ``{Twisting asymptotic symmetries and algebraically special
  vacuum solutions},'' \href{http://dx.doi.org/10.1007/JHEP03(2024)166}{{\em
  JHEP} {\bfseries 03} (2024) 166},
  \href{http://arxiv.org/abs/2401.12054}{{\ttfamily arXiv:2401.12054 [gr-qc]}}.

\bibitem{Townsend:1997ku}
P.~K. Townsend, ``{Black holes: Lecture notes},''
  \href{http://arxiv.org/abs/gr-qc/9707012}{{\ttfamily arXiv:gr-qc/9707012}}.

\bibitem{Barnich:2016lyg}
G.~Barnich and C.~Troessaert, ``{Finite BMS transformations},''
  \href{http://dx.doi.org/10.1007/JHEP03(2016)167}{{\em JHEP} {\bfseries 03}
  (2016) 167}, \href{http://arxiv.org/abs/1601.04090}{{\ttfamily
  arXiv:1601.04090 [gr-qc]}}.

\bibitem{Barnich:2009se}
G.~Barnich and C.~Troessaert, ``{Symmetries of asymptotically flat 4
  dimensional spacetimes at null infinity revisited},''
  \href{http://dx.doi.org/10.1103/PhysRevLett.105.111103}{{\em Phys. Rev.
  Lett.} {\bfseries 105} (2010) 111103},
  \href{http://arxiv.org/abs/0909.2617}{{\ttfamily arXiv:0909.2617 [gr-qc]}}.

\bibitem{Barnich:2010eb}
G.~Barnich and C.~Troessaert, ``{Aspects of the BMS/CFT correspondence},''
  \href{http://dx.doi.org/10.1007/JHEP05(2010)062}{{\em JHEP} {\bfseries 05}
  (2010) 062}, \href{http://arxiv.org/abs/1001.1541}{{\ttfamily arXiv:1001.1541
  [hep-th]}}.

\bibitem{Barnich:2019vzx}
G.~Barnich, P.~Mao, and R.~Ruzziconi, ``{BMS current algebra in the context of
  the Newman\textendash{}Penrose formalism},''
  \href{http://dx.doi.org/10.1088/1361-6382/ab7c01}{{\em Class. Quant. Grav.}
  {\bfseries 37} no.~9, (2020) 095010},
  \href{http://arxiv.org/abs/1910.14588}{{\ttfamily arXiv:1910.14588 [gr-qc]}}.

\end{thebibliography}

\end{document}